**Potential outcome approach to causal inference in assessing the short term impact of air pollution on mortality**

Michela Baccini[1], Alessandra Mattei[1], Fabrizia Mealli[1], Pier Alberto Bertazzi[2,3] Michele Carugno[2]

[1] Dipartimento di Statistica, Informatica e Applicazioni, Università di Firenze, Florence, Italy

[2] Department of Clinical Sciences and Community Health, Università degli Studi di Milano, Milan, Italy

[3] Epidemiology Unit, Department of Preventive Medicine, Fondazione IRCCS Ca' Granda - Ospedale Maggiore Policlinico, Milan, Italy

The opportunity to assess short term impact of air pollution relies on the causal interpretation of the exposure-outcome association, but up to now few studies explicitly faced this issue within a causal inference framework. In this paper, we reformulated the problem of assessing the short term impact of air pollution on health using the potential outcome approach to causal inference. We focused on the impact of high daily levels of particulate matter ≤10 μm in diameter ($PM_{10}$) on mortality within two days from the exposure in the metropolitan area of Milan (Italy), during the period 2003–2006. After defining the number of attributable deaths in terms of difference between potential outcomes, we used the estimated propensity score to match each high exposure-day with a day with similar background characteristics but lower $PM_{10}$ level. Then, we estimated the impact by comparing mortality between matched days. We found that during the study period daily exposures larger than 40 μg/m$^3$ were responsible of 1079 deaths (90% Confidence Interval: 116; 2042). The impact was more evident among the elderly than in the younger classes of age. The propensity score matching turned out to be an appealing method to assess historical impacts in this field.

**INTRODUCTION**

Since the year 2000, many epidemiological studies quantified short term and long term impacts of air pollution on health in terms of the number of sanitary health events due to air pollutant exposures exceeding pre-fixed thresholds (1-5). Short term impact, i.e. the impact observed within few days from the exposure, provides only a partial picture of the health damage attributable to air pollution, because it does not consider consequences of long term exposures, that are characterized by much stronger associations (6-8). However, assessing short term impact has the advantage of allowing an appraisal of the air pollution effect that is not affected by issues that are critical in long term evaluation, such as latency time definition and cumulative exposure assessment (9). Also, short term impact stresses the beneficial effect of measures targeted to immediately improve air quality.

The standard approach to estimate the short term impact of air pollution relies on regression methods. Focusing on mortality, first, the curve describing the relationship between daily exposure and daily deaths is estimated, adjusting for possible confounders, through a regression model; then, the estimated curve is used to calculate how many of the observed deaths are attributable to the exposure levels exceeding a threshold. Varying the threshold, different hypothetical scenarios of air pollution reduction are defined: we can quantify the impact due to exceeding national or international air quality standards, or limits recommended by agencies for public health protection (10).

The opportunity to assess the short term impact of air pollution relies on the causal interpretation of the exposure-outcome association. Up to now, this causal interpretation has been mainly supported by the fact that studies carried out in different countries and contexts provided consistent findings. Moreover, especially for airborne particulate matter, the evidence on the biological mechanisms tying exposure and health damage is consolidated, substantiating the plausibility of the observed associations (11). Bellini et al. (12) read short term effect estimates in light of the Bradford Hill causation criteria (13), showing that they were largely fulfilled. However, also Hill made it explicit that decisions about cause-effect relations cannot be based on a set of rules (14). The principal limitation of this reasoning is the lack of a formal and rigorous definition of causal effect and of the explicit definition of the assumptions needed for a causal interpretation of the epidemiological evidence (15-17).

The potential outcome approach to causal inference, commonly referred to as the Rubin's Causal Model (RCM) (18,19), encourages thinking in terms of causes and action's consequences, within a formal mathematical framework. Despite it is increasingly popular in many fields, including epidemiology and medical sciences, to the best of our knowledge it is relatively new in studies aimed at assessing the impact of air pollution on health. Wang et al. (20) addressed confounding adjustment in model-based estimation of the exposure-response relationship, arguing that their approach is related to causal inference although they do not take a causal inference perspective. In a short commentary to their paper, Gutman and Rubin (21) suggest to use the RCM to estimate the causal effect of air pollution. However, they provide only a theoretical scheme for inference, without any example on real data. More recently, Zigler and Dominici (22) discussed the potential contribution of the potential outcome approach in the policy debate about air pollution regulatory interventions but they did not conduct any empirical analysis. An attempt to use the potential outcome approach to assess causal effects of air pollution on mortality can be found in Schwartz et al. (23) who, however, focus on the effect of a continuous exposure variable using a methodological framework that is different from the one we propose here.

In this paper, we reformulated the problem of assessing the short term impact of air pollution on health within the potential outcome approach to causal inference. In order to illustrate the proposed approach, we assessed the impact of high daily levels of particulate matter ≤10 μm in diameter ($PM_{10}$) on mortality in the metropolitan area of Milan (Italy), during the period 2003–2006. An impact evaluation on the same city and period has been previously conducted following a standard procedure by Baccini et al. (10).

## METHODS

### Data

We considered data for the city of Milan for the years 2003–2006. Milan (1,299,633 inhabitants in 2007) is the capital city of the Lombardy region, in the northwestern Italy. It is located in the basin of the Po River, an area characterized by unfavorable geographical and climate conditions which induce frequent phenomena of thermal inversion. As a consequence, air pollution, mainly deriving from road transport, is trapped close to the ground, reaching very high daily concentrations.

The air quality monitoring network of the Regional Agency of Environmental Protection provided daily measurements of $PM_{10}$, temperature, and relative humidity in the city. A unique daily time series of $PM_{10}$ levels was obtained by averaging data over the available monitors (10). According to large part of the literature, there exists an immediate effect of exposure on mortality which vanishes in few days. Therefore, also in order to allow comparison with previous results, we used the average of the current-day and previous-day $PM_{10}$ concentrations (lag 0-1) as exposure indicator.

Death certificates were obtained from the Regional Mortality Register. We focused on deaths of the resident population occurring inside the city area. We considered daily mortality from all causes except external causes (International Classification of Diseases, Ninth Revision, codes below 800), and, separately, mortality by cardiovascular diseases (ICD-9: 390–459) and respiratory diseases (ICD-9: 460–519). Daily mortality counts were classified by age groups: 15–64 years, 65–74 years, ≥75 years.

### Notation

Indicating with $X_i$ the lag 0-1 exposure in day $i$, $i = 1,...,N$, we defined the treatment indicator $W_i$, equal to 1 if $X_i \geq 40$ µg/m³ (high exposure level) and zero otherwise (low exposure level). Fixing the exposure threshold to 40 µg/m³ assured that the resulting counterfactual time series largely respected the limit of 40 µg/m³ for the annual average concentration of $PM_{10}$, which defines the legal obligation for the European Union member states (24). Then, according to the RCM, under the Stable Unit Treatment Value Assumption (SUTVA) (19), we associated to each day two potential outcomes: $Y_i(1)$, the number of deaths in $i$ if exposure in $i$ was $\geq 40$ µg/m³, and $Y_i(0)$, the number of deaths in $i$ if exposure in $i$ was $< 40$ µg/m³. Obviously, we could only observe at most one of these potential outcomes for each day. Let $Y_i^{obs}$ denote the observed count of deaths in $i$: $Y_i^{obs} = Y_i(0)$ if $W_i = 0$, and $Y_i^{obs} = Y_i(1)$ if $W_i = 1$. We refer to days with $W_i = 1$ as "treated days" and to days with $W_i = 0$ as "control days".

**Definition of attributable deaths**

For each $i$, we defined the day-level AD as the difference between the two potential outcomes:

$$AD_i = Y_i(1) - Y_i(0). \qquad [1]$$

Since we were interested in the total impact of the exposures $\geq 40$ µg/m³ observed during the study period, we focused only on the treated days and defined the total number of AD during the study period as the sum of the day-level impacts in equation 1 for $W_i=1$:

$$AD = \sum_i W_i \left( Y_i(1) - Y_i(0) \right) = \sum_i W_i AD_i. \qquad [2]$$

Being $Y_i(0)$ always missing in equation 2, in order to estimate *AD* we applied a matching procedure to impute these missing potential outcomes: for each treated day $i$, we found one control day with similar background characteristics (*matched control day*), and we used the mortality level observed in this day to impute $Y_i(0)$. We based our matching procedure on the propensity score.

**Design phase: propensity score matching**

Using a matching procedure requires the definition of a distance measure between units. A convenient distance measure, especially when the number of covariates is high, is based on the propensity score (25). Let $Z_i$ a vector of background variables for day $i$. We defined the propensity score as the day-level probability of observing an exposure $\geq 40$ µg/m³, conditional on $Z_i$:

$$e_i = e(\mathbf{Z}_i) = P(W_i = 1|\mathbf{Z}_i). \tag{3}$$

According to Rosenbaum and Rubin (25), if there are no unobserved confounders (unconfoundedness condition) and if there is sufficient overlap in the distribution of the covariates between treated and control days (this assures that for each treated day a controls day with similar background characteristics can be found), adjusting or matching for the propensity score is sufficient for removing confounding. The two conditions mentioned above define the strong ignorability assumption (Appendix, Section S1). A critical issue in the design phase of an observational study is thus the choice of the background variables $\mathbf{Z}_i$ conditionally on which strong ignorability is reasonable. We based this selection on a priori substantive knowledge of the phenomenon deriving from the literature on the short term effects of air pollution, which suggests that the air pollution-mortality relationship can be confounded by meteorological conditions, short term and long terms seasonality and other factors that could produce unusual picks of mortality.

*Propensity score estimation*

The propensity score for each unit was estimated from a logistic model for $W_i$, including terms for all the relevant background variables $\mathbf{Z}_i$:

$$W_i \sim Bernoulli(e_i) \quad \text{logit}(e_i) = f(\mathbf{Z}_i, \boldsymbol{\beta}), \tag{4}$$

where $\boldsymbol{\beta}$ was a vector of unknown coefficients and $f$ a general function of covariates and coefficients. Then, indicating with $\hat{\boldsymbol{\beta}}$ the vector of the estimated coefficients, the estimated propensity score was obtained as:

$$\hat{e}_i = \frac{\exp(f(\mathbf{Z}_i, \hat{\boldsymbol{\beta}}))}{1 + \exp(f(\mathbf{Z}_i, \hat{\boldsymbol{\beta}}))}. \tag{5}$$

Different specification were possible for the model in equation 4 and some effort was needed to find an appropriate $f$. Being the propensity score a balancing score (25), the key criterion driving the specification of $f$ consists in obtaining predicted values $\hat{e}_i$ conditionally on which the covariates distribution is the same in the treated and matched control groups.

We assessed the balancing property for each covariate, under different choices of $f$, by using various diagnostic tools: visual inspection of the distributions before and after matching, comparison of pre- and post-matching standardized mean differences (when applicable) (Appendix, Section S2), statistical tests for pre and post-matching comparisons of means and proportions. Regarding seasonality, we assessed balance by comparing the median month between groups by using a nonparametric test, which accounted for the circular nature of the month-variable (26). The model specification that led to the best balance in covariates distributions included season-specific indicators of day of week and holiday, an indicator of days with influenza epidemics, a cubic regression spline with 5 degrees of freedom per year on the calendar day to account for medium and long term seasonality, and a bivariate smooth term for temperature at lag 0-3 and humidity, defined by the tensor product of two marginal thin plate regression splines with basis dimensions 5 and 3, respectively (27). We included in the model also an indicator of days with temperature exceeding 28°C to capture possible effect of extreme heat episodes, and an indicator of the July-August period to account for the reduction of the population present in the city during summer holidays.

*Nearest neighbor matching*

For each treated day *i*, we selected as match the control day with estimated propensity score closest to *i* (*nearest neighbor matching*) (28). We used matching with replacement, allowing each control day to be used as a match more than once, because matching with replacement produces matches of higher quality than matching without replacement, thus reducing bias even at the cost of some precision (29).

**Analysis phase: AD estimation**

For each treated day *i*, we first imputed the missing potential outcome $Y_i(0)$ using the count of deaths observed in its matched-control day $Y_i^C$. Then, we estimated the day-level impact for each treated day as the difference:

$$\widehat{AD}_i = Y_i^{obs} - Y_i^C. \quad [6]$$

Finally, we estimated *AD*:

$$\widehat{AD} = \sum_i W_i \widehat{AD}_i. \quad [7]$$

The estimate of the variance of $\widehat{AD}$ was derived from the sample variance reported in Abadie and Imbens (29) for the Sample Average effect of Treatment on the Treated estimator (Appendix, Section S3).

All analyses were performed in R software (R Core Team; http://www.R-project.org/).

**RESULTS**

On average, in Milan between 2003 and 2006 there were 31.2 natural deaths per day, 10.3 from cardiovascular causes and 2.5 from respiratory causes. The annual average level of exposure ($PM_{10}$ at lag 0-1) was 52.5 μg/m$^3$; 812 days during the study period (55.7%) exceeded 40 µg/m$^3$ and 593 days (40.7%) exceeded the EU daily limit of 50 µg/m$^3$ (24); the exposure level was sometimes very high, with around 9% of days exceeding 100 µg/m$^3$.

Table 1 and Figure 1 report the results of the balancing property check for the selected propensity score model. While the propensity score distributions in treated and control days were very different, the distribution in matched control days was completely overlapped with the distribution in treated days. The treated days were characterized by very different temperature and relative humidity with respect to the control days. However, the distributions of the meteorological variables in the matched samples were similar, as confirmed by the t tests, which did not reject the null hypothesis of equal means, and by the standardized differences, which were very small after matching. Regarding the percentage of heat episodes, it was similar among treated and controls both before and after matching. Days characterized by peaks of influenza were 12.8% in the treated group and only 0.9% in the control group. After matching, the standardized difference reduced by 34.7%, but balancing remained unsatisfactory, with a very small p-value for the test of the equality between proportions. This was not completely unexpected because influenza epidemics may affect the outcome by acting jointly to the exposure, and a matching procedure based on propensity score could be not fully successful in balancing its effect. In order to check robustness of our results, we performed a sensitivity analysis, which investigated the impact on the subset of days without influenza epidemics, by simply excluding days with influenza epidemics (see below).

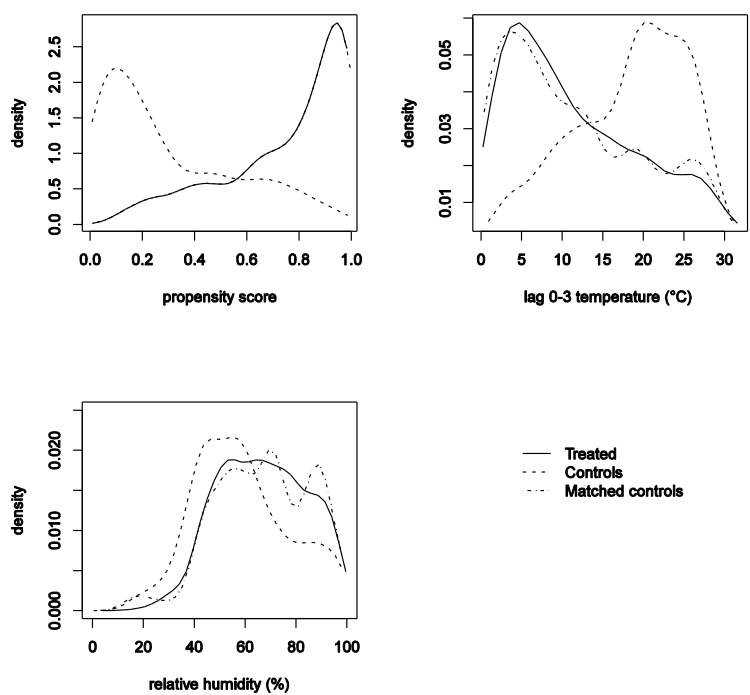

**Figure 1.** Density functions of estimated propensity score, average temperature in the current and in the previous three days (lag 0-3) and relative humidity for treated days, control days and matched control days, Milan, Italy, 2003-2006.

**Table 1.** Covariates Balance Before and After Matching, Milan, Italy, 2003-2006.

| Background Characteristic | Mean/Proportion | | | P | | Standardized Difference[d] | | % Bias[e] |
|---|---|---|---|---|---|---|---|---|
| | Treated | Controls | Matched Controls | Pre-matching | Post-matching | Pre-matching | Post-matching | |
| Estimated Propensity Score | 0.756 | 0.306 | 0.756 | <0.001 | 1 | 1.810 | 0 | 100.0 |
| Temperature (°C)[a] | 11.4 | 18.3 | 11.3 | <0.001 | 0.813 | 0.914 | 0.013 | 98.5 |
| Relative Humidity (%) | 66.8 | 58.6 | 67.1 | <0.001 | 0.780 | 0.456 | 0.014 | 97.0 |
| Saturdays and Sunday | 0.243 | 0.341 | 0.195 | <0.001 | 0.950 | 0.217 | 0.106 | 51.0 |
| Calendar Month | | | | <0.001[f] | 0.322[f] | | | |
| Influenza Epidemics | 0.128 | 0.009 | 0.054 | <0.001 | <0.001 | 0.483 | 0.315 | 34.7 |
| Heat Episodes[b] | 0.032 | 0.028 | 0.025 | 0.759 | 0.454 | 0.001 | 0.002 | -77.8 |
| Summer Days | 0.225 | 0.664 | 0.252 | <0.001 | 0.222 | 0.037 | 0.002 | 93.8 |

P: P-value.
[a] Temperature: average temperature in the current and in the previous 3 days.
[b] Heat episodes: days with temperatures exceeding 28°C. Summer days: from May 1st to September 30th.
[c] p-value: p-values of the test for the comparison of means or proportions between treated and controls (pre-matching p-values) or between treated and matched controls (post-matching p-values).
[d] Standardized difference: pre-matching ($\delta^{pre}$) and post-matching ($\delta^{post}$) (see Online Resource, Section S1).
[e] % bias: $100 \times (\delta^{pre} - \delta^{post})/\delta^{pre}$.
[f] p-value of the test for the difference between circular medians.

In order to check balance for day of week, we focused on the percentage of Saturdays and Sundays, which resulted very similar in treated and matched control days. In order to check seasonality balance, we focused on a warm season indicator, which was equal to 1 from May 1st to September 30th and 0 elsewhere, and on calendar month. Matching clearly reduced the percentage of summer days in favor of winter days (Figure 2). Substantial balance after matching was found for the percentage of warm season days, with p value after matching equal to 0.222 and % bias reduction equal to 93.8%. Also calendar month resulted well balanced after matching: the test on the circular medians of the calendar month distribution, which was highly significant before matching, did not reject the null hypothesis of equivalence after matching (P=0.32).

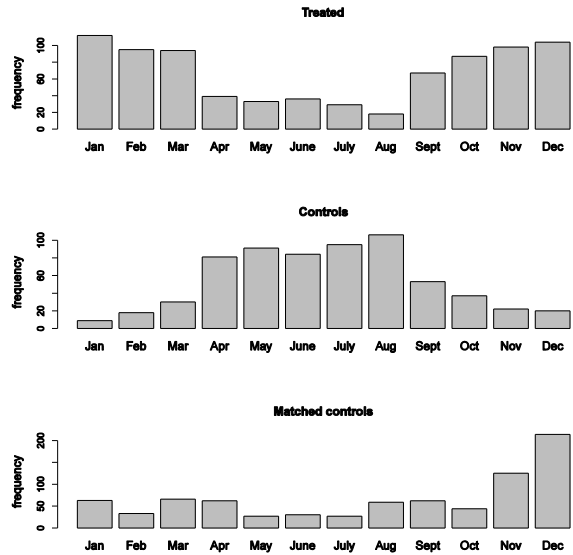

**Figure 2.** Distributions of treated days, control days and matched control days by calendar month, Milan, Italy, 2003-2006.

Figure 3 shows the daily number of natural deaths during the study period, together with daily exposures and daily-level impacts by calendar day. Day-level impacts appeared to be rather heterogeneous. Exposures ≥40 μg/m$^3$ were observed mainly during winter, but we estimated relevant positive day-level impacts also during summer, which could be indicative of a possible an interaction between temperature and exposure. We could be floored by the presence of negative estimates of the day-level impacts (Figure 3). This might lead to misleading conclusions if we do not consider that each $\widehat{AD}_i$ depends on the imputed value of the missing potential outcome $Y_i(0)$. This imputation is naturally subject to variability, which may lead to negative estimated impacts, even for days where air pollution had indeed small or no impact. In this sense, negative $\widehat{AD}_i$ are fully consistent with the existence of a dangerous effect of the exposure.

**Table 2**. Estimated Number of Attributable Deaths by Cause and Age Class, Milan, Italy, 2003-2006.

|  | Age 15-64 | | Age 65-74 | | Age 75+ | | All ages (15+) | |
| --- | --- | --- | --- | --- | --- | --- | --- | --- |
|  | AD | 90% CI | AD | 90% CI | AD | 90% CI | AD | 90% CI |
| **Cardiovascular Causes** | -172 | -368, 24 | 91 | -244, 426 | 797 | 305, 1288 | 716 | 117, 1315 |
| **Respiratory Causes** | -25 | -133, 83 | 87 | 11, 163 | 243 | -22, 508 | 305 | 17, 593 |
| **Other Natural Causes** | 153 | -246, 552 | -157 | -401, 87 | 62 | -414, 538 | 58 | -496, 612 |
| **All Natural Causes** | -44 | -609, 521 | 21 | -425, 467 | 1102 | 388, 1816 | 1079 | 116, 2042 |

AD: Attributable deaths; 90% CI: 90% Confidence Interval.

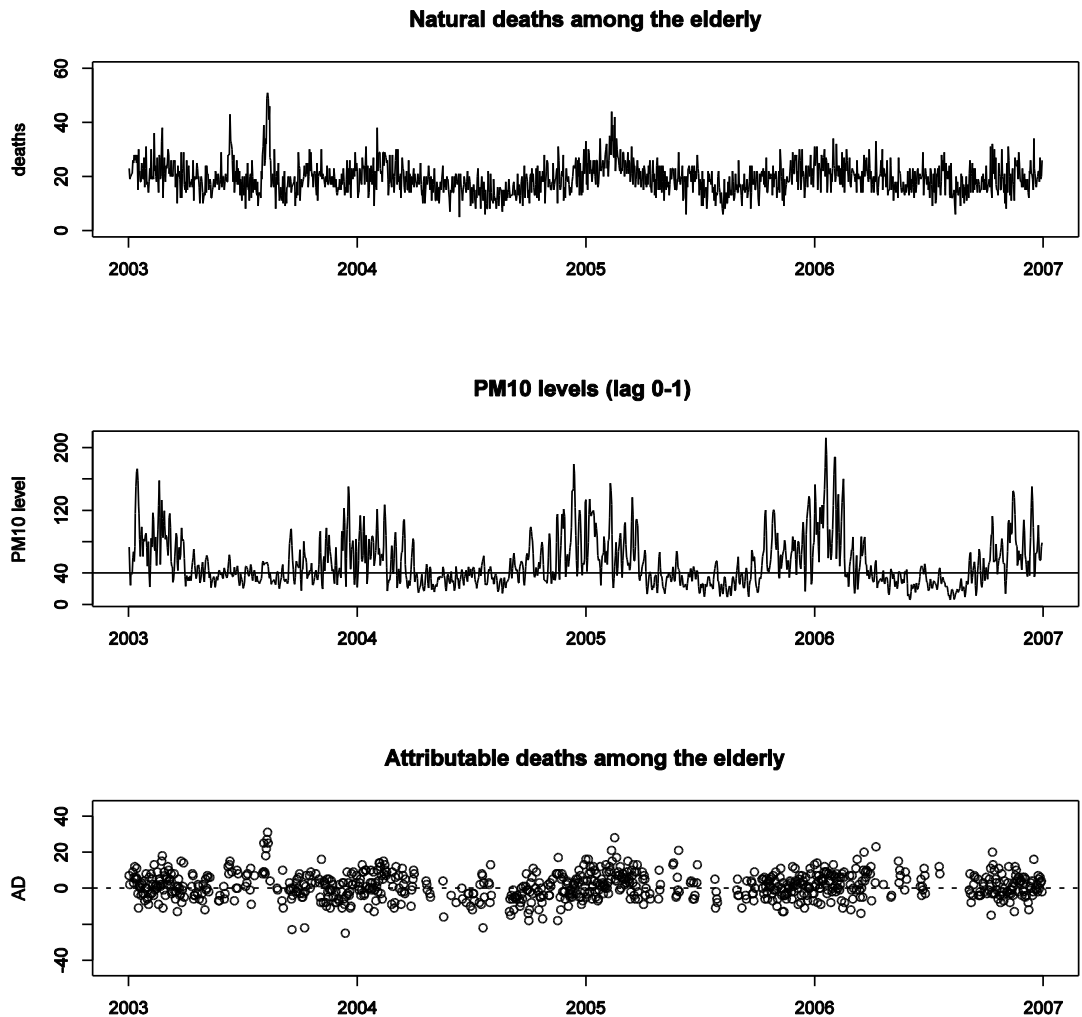

**Figure 3.** Daily counts of deaths among people aged 75 and over (upper panel), average $PM_{10}$ level in the current and in the previous day (lag 0-1) (middle panel) and estimated daily attributable deaths (lower panel), Milan, Italy, 2003-2006.

Table 2 shows the estimated AD for each cause of death and class of age, along with their 90% confidence intervals. These results should be interpreted as the number of deaths that would have been saved on average, if the daily level of exposure had never exceeded 40 μg/m³ during the study period. The impact was concentrated among individuals over 75. Exposures ≥ 40 μg/m³ were responsible of 1102 AD among the elderly (90% CI: 388, 1816), 797 of which for cardiovascular causes (90% CI: 305, 1288) and 243 for respiratory causes (90% CI: -22, 508); the impact on mortality from natural causes except respiratory or cardiovascular ones was much lower (62 AD) and affected by large uncertainty. Clear evidence of an impact of air pollution on respiratory mortality was found also in the age class 65-74, with 87 AD (90% CI: 11, 163). For the first class of age (15-64) the confidence intervals were extremely wide, so that clear conclusions could not be drawn. It is worth noticing that the total estimated AD (1079) corresponded to the sum of the age- and cause-specific impacts; the confidence interval around this value was large, but clearly far from zero (90% CI: 116, 2042).

Excluding control days which have never been selected as matched controls (430), 46% of the remaining control days were selected once, 84% less than 5 times, 95% less than 15 times. Few days were selected a very large number of times (Figure 4). The main consequence of using the same control day as match many times is an increase of the estimated variance, although with a benefit in terms of bias (29). In order to get some insight on the influence of control days used as matches multiple times on the impact estimates, we investigated mortality and air pollution levels among those days. We did not find any influent point; the control day that was used as match for more than 60 treated days was characterized by a large number of deaths, thereby leading to a possible conservative lower impact.

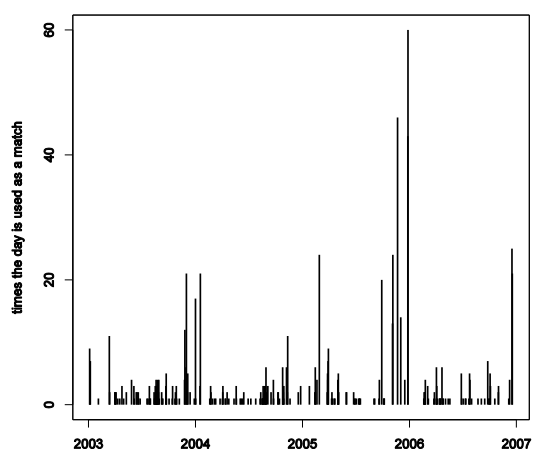

**Figure 4.** Number of times each control day is selected as a matched control, Milan, Italy, 2003-2006.

**Table 3.** Estimated Number of Attributable Deaths by Cause and Age Class, After Excluding Influenza Epidemic Days, Milan, Italy, 2003-2006.

|  | Age 15-64 | | Age 65-74 | | Age 75+ | | All ages (15+) | |
|---|---|---|---|---|---|---|---|---|
|  | AD | 90%CI | AD | 90%CI | AD | 90%CI | AD | 90%CI |
| **Cardiovascular Causes** | -78 | -219, 63 | 93 | -108, 294 | 469 | 88, 850 | 484 | 12, 956 |
| **Respiratory Causes** | -22 | -91, 47 | 57 | -7, 121 | 276 | 99, 452 | 311 | 122, 500 |
| **Other Natural Causes** | 28 | -257, 313 | 16 | -248, 280 | -38 | -434, 358 | 6 | -514, 526 |
| **All Natural Causes** | -72 | -456, 312 | 166 | -182, 513 | 707 | 100, 1314 | 801 | 6, 1595 |

AD: Attributable deaths; 90% CI: 90% Confidence Interval.

Table 3 shows the results of the sensitivity analysis performed excluding from AD calculation the days of influenza epidemics. While AD for respiratory causes were substantially unchanged, the impact estimate on cardiovascular mortality among the elderly were lower, although still relevant (469 AD, 90% CI: 88, 850). We can conclude that our results were robust to possible bias derived from the confounding effect due to influenza epidemics.

**DISCUSSION**

Our analysis confirms that restraining PM10 exposure under the EU limits could have saved a relevant number of deaths in Milan during the study period, with an estimated impact even larger the impact reported in Baccini et al. (10) following an approach based on Poisson regression. While we estimated a total of 1079 AD, Baccini et al. (10) found that exceeding the limits of 40 and 20 μg/m$^3$ for the annual average concentration of $PM_{10}$ was responsible of 358 and 925 deaths for natural causes (89.5 and 231.3 per year), respectively. However, for a fair comparison, we need to consider that in our analysis the counterfactual annual average exposure was lower than 40 μg/m$^3$ (being the 40 μg/m$^3$ counterfactual level defined on the daily exposures), and that the impact in Baccini et al. (10) was based on a shrunken estimate of the exposure-mortality association arising from a meta-analysis, which was, at least partially, influenced by the lower effect estimated at the regional level.

Our approach relies on well defined assumptions. SUTVA requires that there are not hidden version of the treatment and that the potential outcomes on one unit are unaffected by the specific treatment assigned to the other units (no-interference among units) (19). In our context, this second condition could be critical, because the exposure in a day could affect not only mortality in the current day, but also in subsequent days. Focusing on the lag 0-1 exposure instead of on the current $PM_{10}$ level makes the no-interference assumption more plausible. Obviously, enlarging the window of the moving average $X_i$ would empower the no-interference assumption, but at the price of a lower variability of the exposures and a reduced possibility of detecting an impact if any. The other relevant assumption is the unconfoundedness assumption (25). Being this condition not directly testable on data, we grounded its plausibility on subject-matter knowledge deriving from the literature.

The idea of using matching is not new in the analysis of the short term effects of air pollution on health. The most popular example of matching in this field is the case-crossover approach, proposed as an alternative to Poisson regression with the aim to adjust for the confounding effect of seasonality by design (30). However, it is worth noting that the rationale of the propensity score-based matching is different from the rationale of the case-crossover approach. In our analysis matching is done on the exposure variable: we matched each high exposure day with the low exposure day exhibiting the closest propensity score. On the contrary, the case-crossover approach matches on the outcome: an individual who died in a certain day is matched with himself in one or more days when he did not die. Moreover, despite the use of matching, the case-crossover approach is much closer to the standard analysis based on Poisson regression than to the approach proposed in this paper, and shares drawbacks and advantages with Poisson regression (31).

The approach we proposed has several advantages over the standard approach based on regression. The first one stems from the fact that it clearly distinguishes between the design phase and the analysis phase. The design phase (from propensity score estimation to matching) does not involve outcome data, but only background information. As a consequence, the sub-sample of units arising from the design phase (treated units with the corresponding matched controls) can be used in the analysis phase to estimate the causal effects of the treatment on one or more outcomes, in our case cause-specific and age-specific mortalities. This also implies that results on different outcomes are fully consistent. For instance, the estimated total number of AD can be derived directly as sum of either age-specific or cause-specific AD. This consistency is not guaranteed within the standard model-based approach, because impact estimation by age and mortality cause is usually derived by fitting separated regression models for each outcome. For these characteristics, the proposed approach is promising to detect susceptible subpopulations and to perform surveillance focusing on very specific causes of death or diseases.

A second advantage concerns the interpretation of the results. Our approach forces us to explicitly define the assumptions needed for draw inference on the causal quantities of interest. On the contrary, results from regression rely on strong assumptions that often are not explicitly stated making causal interpretation of the results controversial. Moreover, by clearly specifying the critical assumptions, we can assess the consequences of their violation; for instance, methods to evaluate robustness of the results to possible violation of unconfoundedness can be applied (see Chapter 21 and 22 in (19) for a comprehensive review).

A third advantage is that our approach is free from issues concerning the exposure-confounders-mortality modeling and does not involve extrapolation. Although this may sacrifice some external validity, implying that inferences (for example attributable fractions) are less likely to be valid for populations with different characteristics from those observed in the sample, it awards a strong internal validity to impact estimates. Problems related to poor model fit and inappropriate extrapolations due to limited overlap in the distributions of the covariates between treated and controls can be detected and addressed more easily using the approach we propose.

In order to correctly interpret the results of our analysis, we also need to account for some critical points. We considered the treatment as binary, after having defined an arbitrary, although substantive, threshold. In this sense our impact estimates are not directly comparable with those arising from the standard approaches, which usually assume a linear relationship between exposure and mortality on a logarithmic scale. A valuable topic for future research is to apply methods for a continuous exposure (32, 33).

We did not fix the daily counterfactual to a specific value: for instance, in our study the air pollution level in the matched control days ranged from the observed minimum exposure to 40 μg/m$^3$.

Our approach allowed us to consider counterfactual scenarios defined on daily exposures, but not in terms of annual average concentration, although these could be of interest from a legal and regulatory standpoint.

Finally, although appealing to assess historical impacts, the proposed method is not appropriate to estimate future impacts. Despite this, it should be considered as a tool to check internal validity of regression-based results, before any use of the estimated associations for projections purposes.

# APPENDIX

## Section S1. Strong ignorability assumption

Let $\mathbf{Z}_i$ denote the vector of the observed covariates for day $i$. The treatment assignment mechanism is strongly ignorable if the following conditions hold (1):

(i) Unconfoundedness: Treatment assignment is independent of the potential outcomes conditional on the observed covariates: $W_i \perp (Y_i(0), Y_i(1)) | \mathbf{Z}_i$ ;

(ii) Overlap: $0 < P(W_i = 1 | \mathbf{Z}_i = \mathbf{z}) < 1$ for each $i$.

Because we are only interested in the impact for treated days, we can weaken the strong ignorability assumption as follows:

(i) Unconfoundedness for controls: Treatment assignment is independent of the potential outcome under control conditional on the observed covariates: $W_i \perp Y_i(0) | \mathbf{Z}_i$ ;

(ii) Weak overlap: $P(W_i = 1 | \mathbf{Z}_i = \mathbf{z}) < 1$ for each $i$.

Unconfoundedness requires that there are no unobserved confounders, so that conditioning on the observed covariates assures that an experimental-like context is reproduced. The overlap assumption requires that there is sufficient overlap in the joint distribution of the covariates between treated and control days. This second assumption is needed because, if all days with given background characteristics had exposure larger than the threshold ($P(W_i = 1 | \mathbf{Z}_i = \mathbf{z}) = 1$ for some $\mathbf{z}$), then there would be no similar control days against which to compare them.

## Section S2. Pre- and post-matching standardized mean differences

For each background variable, we calculated the pre- and post-matching standardized mean differences which are defined as follows:

$$\delta^{pre} = \frac{\bar{x}_{treated} - \bar{x}_{control}}{\sqrt{\frac{s^2_{treated} + s^2_{control}}{2}}} \quad \text{and} \quad \delta^{post} = \frac{\bar{x}_{treated} - \bar{x}_{matched}}{\sqrt{\frac{s^2_{treated} + s^2_{control}}{2}}} \qquad [1]$$

where $\bar{x}_{treated}$, $\bar{x}_{control}$ and $\bar{x}_{matched}$ are the sample averages of the covariate for controls, treated and matched controls, and $s^2_{treated}$ and $s^2_{control}$ are the within-group sample variances for treated and controls.

## Section S3. Sample variance of $\widehat{AD}$

The estimate of the variance of $\widehat{AD}$ was derived from the sample variance proposed by Abadie and Imbens (2,3) for the estimate of the Sample Average effect of Treatment on the Treated (SATT) estimator:

$$s^2_{AD} = \sum_{i=1}^{N} \left(W_i - (1-W_i)K_M(i)\right)^2 \sigma^2_{W_i}(\mathbf{Z_i}), \qquad (2)$$

where $K_M(i)$ is the number of times the unit $i$ is used as a match, and $\sigma^2_{W_i}(\mathbf{X_i})$ is the conditional variance of the outcome. The conditional variance can be estimated accounting for possible heterogeneity according to the following equation:

$$\sigma^2_{W_i}(\mathbf{X_i}) = \frac{1}{\#H_M(i)} \sum_{j \in \{H_M(i) \bigcup (i)\}} \left(Y_j - \bar{Y}_{H_M(i) \bigcup (i)}\right)^2, \qquad [3]$$

where $H_M(i)$ is the set of indexes for the first $M$ matches for unit $i$, $\#H_M(i)$ is the number of elements of $H_M(i)$ and $\bar{Y}_{H_M(i) \bigcup (i)}$ is the mean of the outcome in the set including the unit $i$ and its matches. It is worth noting that the variance estimator (2) does not account for the fact that the propensity score is estimated. Equations 2 and 3 are valid for each $M \geq 1$; in our analysis we set $M = 1$.